# Overcoming linguistic barriers in code assistants: creating a QLoRA adapter to improve support for Russian-language code writing instructions

**Pronin C.B.[1], Volosova A.V.[2], Ostroukh A.V.[1], Strogov Y.N.[1]**

[1] Moscow Automobile and Road Engineering State Technical University (MADI), Leningradsky Prospekt, 64, 125319 Moscow, Russia.

[2] Bauman Moscow State Technical University, 2nd Bauman str., 5, p. 105005 Moscow, Russia.

**Abstract**: In this paper, an approach to training and evaluating an adapter model for the popular language model "zephyr-7b-beta" is described. The adapter was developed to improve the performance of the base model in tasks related to programming and understanding the Russian language. Considering the high quality of the original model in tasks in the English language, the goal of the research was to expand its linguistic and technical spectrum. The proposed adapter was trained using a large and diverse dataset, including question-answer pairs related to programming, as well code-related texts in Russian language. The applied training methodology ensures an improvement in the model's quality of answers in understanding and generating Python code based on Russian instructions. We evaluated the performance of the base model with the installed adapter using various metrics, comparing it to the base model as well as other state-of-the-art models in this field. The obtained results showed significant improvement, both in tasks related to writing Python code and in processing the Russian language, confirming the effectiveness of the proposed adapter.

# Introduction

The language model 'zephyr-7b-beta' stands out for its high quality performance in processing English texts. Therefore, expanding its linguistic and technical scope is quite interesting. In this regard, an approach to training and evaluating the adapter model for this popular language model is proposed.

The goal of the experiment conducted in this paper is to fine-tune the original large language model in order to improve its capabilities in writing Python code based on Russian instructions and providing explanations in Russian.

The content of the basic dataset consists of 49.8 thousand lines of synthetic programming instructions translated by machine translation, and 15.1 thousand lines of discussion of Python code in natural language. Before training, the instruction set underwent formatting and deduplication.

The experiment is based on the HuggingFaceH4/zephyr-7b-beta model, which has shown extremely high test results in text processing, but is slightly inferior in code writing and solving mathematical problems compared to "proprietary" language models (Fig. 1) [1]. This model has already been fine-tuned on a large and diverse set of instructions.

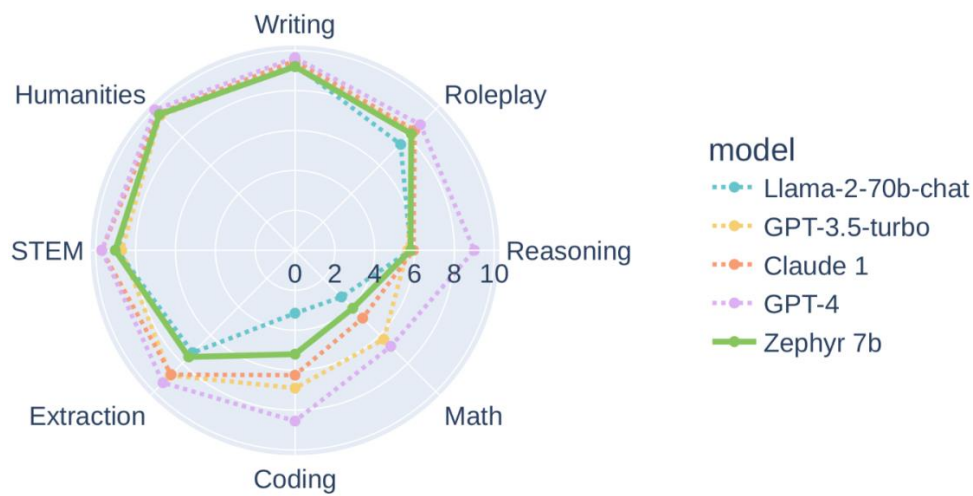

*Fig. 1. Performance of base model ZEPHYR-7B on MT-Bench [1]*

Fine-tuning language models using LoRA (Layer-wise Relevance Approximation) and QLoRA (Quantized LoRA) offers several advantages. These methods are part of the

PEFT (Parameter Efficient Fine-Tuning) library, which significantly reduces the number of trainable parameters without visible quality degradation, thus allowing even the largest language models to be fine-tuned with significantly fewer computational resources [2, 3].

Since the training of the adapter model (LoRA/QLoRA) does not modify the weights of the base model (except for possible quantization), there is a common belief that this method of fine-tuning neural network models can avoid catastrophic forgetting of the knowledge learned during the pre-training phase. Therefore, it is reasonable to assume that the original model can be further fine-tuned with new instructions, on top of the instructions already incorporated during "instruction fine-tuning," in order to improve the model's performance in specific tasks without losing its original behavior.

The results of the developed model's testing [4] showed that the model's behavior changed significantly compared to the behavior of the base model, favoring code generation for solving the given task in response to any short instruction. However, the response to longer instructions remains at the level of the base model. The results of synthetic tests showed:

Perplexity lower than that of the base model (lower perplexity indicates better performance).

Accuracy in solving GSM8K mathematical problems higher than that of the base model, with some deterioration in performance in other testing categories.

Based on the testing results, it can be concluded that the adapter model can have different effects (both positive and negative, depending on the training parameters and dataset) on the model's performance beyond the set of tasks defined in the training data.

**Testing the trained adapter model**

Perplexity is one of the most common metrics for evaluating the quality of language models. A language model (in the context of NLP) refers to the probability distribution over a set of sequences (tokens[1]). As an example, let's consider an n-gram sequence for n=2. Applying a 2-gram model to the sentence "Speech recognition is undoubtedly a challenging task," ("Распознавание речи несомненно является сложной задачей") we obtain [5]:

$$P(\text{Распознавание, речи, несомненно, является, сложной задачей} \approx P(\text{Распознавание} | <s>) \times P(\text{Речи} | \text{Распознавание}) \times P(\text{несомненно} | \text{речи}) \times P(\text{является} | \text{несомненно}) \times P(\text{сложной} | \text{является}) \times P(\text{задачей} | \text{сложной}) \times P(</s> | \text{задачей}) \quad (1)$$

This metric can be considered as a measure of uncertainty when predicting the next sequence in a language model.

Traditionally, entropy is used as a measure of uncertainty. The emergence of perplexity as another metric related to uncertainty in a language model is due to the need to minimize the degree of this uncertainty. This is a rather challenging process, as calculating the optimal value of uncertainty for a specific model is a complex task. If we consider language entropy as:

$$H = \lim_{N \to \infty} F_n, \quad (2)$$

where $F_N$ – function, describing information quantity, related to $N$ adjacent tokens of text $T$:

$$F_N = -\sum_{cw_n} p(cw_n) \log_2 p(cw_n) + \sum_{cw_{n-1}} p(cw_{n-1}) \log_2 p(cw_{n-1}), \quad (3)$$

where $cw_n$ – corpus of n tokens, adjacent with the text.

Based on the given text, a language model can determine a distribution $D_T$ that is close to the empirical distribution of the language - $D_L$. Such an approximation is related to an infinite number of sequences considered in (2). To determine the degree of similarity between $D_L$ and $D_T$, a metric called cross-entropy is used:

$$H(D_L, D_T) = E_p[-\log D_T] \quad (4)$$

with discrete $D_L$ and $D_T$ (3) takes form:

$$H(D_L, D_T) = H(D_L) + D_{KL}(D_L || D_T), \quad (5)$$

where $D_{KL}$ – is the Kullback-Leibler divergence.

---
[1] By sequence we mean a symbol, a word or part of a word

Thus, the task of optimizing the degree of uncertainty boils down to minimizing the Kullback-Leibler divergence, which is available to the language model as a result of studying $D_L$.

For the convenience of evaluating the language model, perplexity is used as another metric that characterizes the degree of uncertainty when predicting the next sequence in the considered language model:

$$Perplexity(D_L, D_T) = 2^{H(D_L, D_T)} \qquad (6)$$

Perplexity allows us to evaluate the complexity of a language model.

For example, if the perplexity has a low value, then we are dealing with a text that consists of grammatically correct sentences. If the perplexity is high, then the text is considered incorrect.

Based on the definition of a language model, to calculate the perplexity of a sentence S composed of K tokens (t), the following formula can be used:

$$Perplexity(S) = \sqrt[K]{\frac{1}{P(t_1,\ t_2,\ ...,\ t_K)}} \qquad (7)$$

Thus, perplexity can be used to evaluate the quality of a language model and can be defined:

1. Using cross-entropy:

$$Perplexity(S) = 2^{H(S)}, \qquad (8)$$

where

$$H(S) = 2^{-\frac{1}{K}\log_2 P(t_1,\ t_2,\ ...,\ t_K)} \qquad (9)$$

Perplexity corresponds to the number of words that can be encoded with the average number of bits (the value of cross-entropy) required to encode one word.

2. As the inverse probability of a text set, normalized by the number of words (see formula 7).

During the experiment, an adapter model (QLoRA) was trained to complement the base model without modifying it. Adapter tests on perplexity are shown in Figures 2 and 3 (lower values corresponding to our results are considered better).

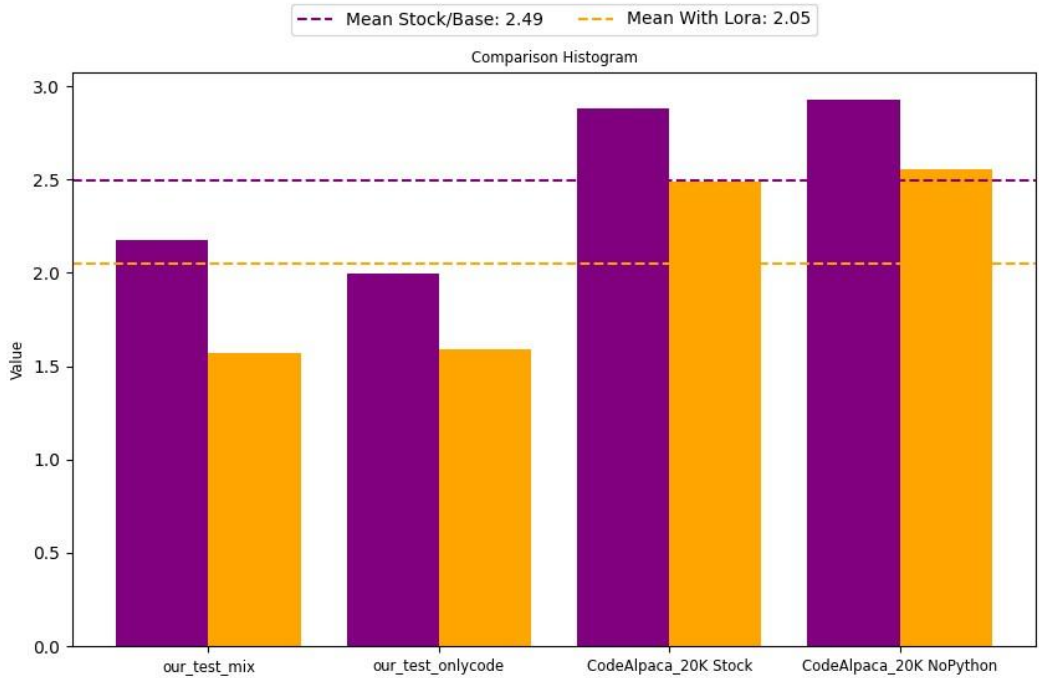

*Fig. 2. Perplexity test (from left to right) on: 1) a test sample from our dataset that is not involved in the model training process; 2) only code and instructions from this dataset; 3) the CodeAlpaca_20K dataset, which did not participate in adapter training [6]; 4) CodeAlpaca_20K without python code.*

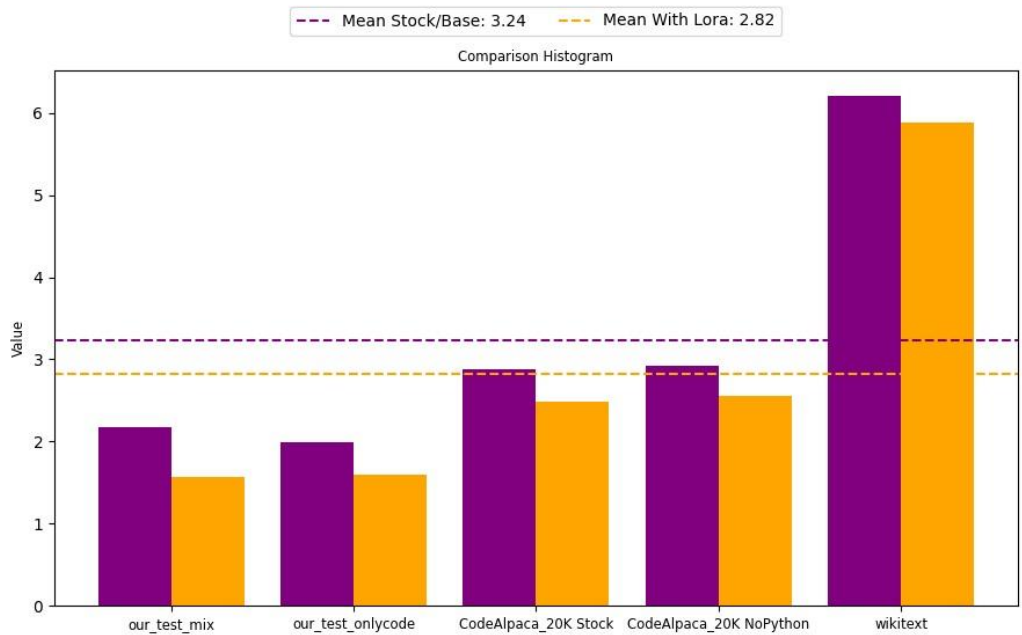

*Fig. 3. Perplexity test results after adding results from the wikitext dataset*

Perplexity is not the only metric applicable for evaluating language models. To compare how the model with the proposed adapter differs from the original one, tests were conducted on the model with the built-in adapter (merged) in an automated language model testing system using multiple metrics - HuggingFace Open LLM Leaderboard (Fig. 4) [7]. For the convenience of model inference, models were created in GGUF format quantized to 4 and 6 bits. Converting the model to GGUF format (fp16) with subsequent quantization (to q4_k_m or q6_k) provides high debugging speed on both CPU and GPU acceleration. In this case, GGUF (using llama.cpp) allows using the random-access memory (RAM) as a buffer for loading parts of the model, enabling GPU acceleration even when there is insufficient video memory to fully load the model (when setting the parameter n_gpu_layers > 0).

| T | Model | Average | ARC | HellaSwag | MMLU | TruthfulQA | Winogrande | GSM8K |
|---|---|---|---|---|---|---|---|---|
| ? | Fredithefish/OpenZephyrChat | 68.12 | 64.85 | 85.08 | 64.92 | 48.24 | 81.06 | 64.59 |
| ◆ | Weyaxi/OpenOrca-Zephyr-7B | 64.97 | 64.08 | 83.82 | 62.46 | 54.31 | 78.93 | 46.25 |
| ◆ | uukuguy/zephyr-7b-alpha-dare-0.85 | 62.35 | 61.18 | 83.67 | 64.3 | 44.41 | 78.45 | 42.08 |
| ◆ | abhishek/zephyr-beta-math | 61.99 | 56.66 | 81.26 | 57.24 | 44.83 | 75.53 | 56.41 |
| ◆ | HuggingFaceH4/zephyr-7b-beta | 61.95 | 62.03 | 84.36 | 61.07 | 57.45 | 77.74 | 29.04 |
| ◆ | HuggingFaceH4/zephyr-7b-beta | 61.59 | 62.46 | 84.35 | 60.7 | 57.83 | 77.11 | 27.07 |
| ○ | tianlinliu0121/zephyr-7b-dpo-full-beta-0.2 | 61.55 | 61.77 | 84.04 | 61.79 | 54.72 | 76.95 | 30.02 |
| ○ | tianlinliu0121/zephyr-7b-dpo-full-beta-0.2 | 61.36 | 61.86 | 83.98 | 61.85 | 54.78 | 76.95 | 28.73 |
| ◆ | Sao10K/Zephyrus-L1-33B | 60.61 | 64.51 | 84.15 | 57.37 | 53.87 | 80.19 | 23.58 |
| ◆ | MexIvanov/zephyr-python-ru-merged | 60.1 | 56.06 | 82.06 | 60.2 | 52.81 | 76.95 | 32.52 |
| ◆ | HuggingFaceH4/zephyr-7b-alpha | 59.5 | 61.01 | 84.04 | 61.39 | 57.9 | 78.61 | 14.03 |
| ○ | mergedlm/zephyrnotus-11b-alpha | 59.26 | 61.35 | 82.8 | 60.67 | 57.22 | 76.4 | 17.13 |

*Fig. 4. HuggingFace Open LLM Leaderboard – top models by search filter "Zephyr" date: 25.12.2023*

## Conclusions

In the process of research, the following results were obtained:

1. During the conducted experiment, an assessment of the possibility of fine-tuning a pre-trained model on a set of instructions was performed. The goal was to achieve additional fine-tuning with new instructions without losing the ability to perform the original instructions. Based on the description of the LoRA method, such an approach is possible because during this type of training, the original model weights remain unchanged and only new weights (as coefficients) are added from the adapter model, which was trained on new instructions.

2. The analysis of the experiment results showed that the adapter not only adds new functionality to the model but significantly changes the priority of model output in favor of the instructions that were included during fine-tuning.

3. To eliminate the fact of changing the priority of model output mentioned above, we propose the following methods:
   3.1. Using a different prompt format when formatting the training data set for the model, different from the original Zephyr format.
   3.2. Including instructions from the first stage of fine-tuning.

3.3. Creating a "mixture of experts" from several different models, where each model will be specialized in solving one or several tasks from the general spectrum of solvable tasks.

The ways (described above) to fine-tune a model without losing its original characteristics are stated as promising directions for future research in the field of resource-efficient modification of large language models.

The presented research contributes to current efforts in machine learning and natural language processing to create more universal AI models. The observed potential of adapter models in improving the quality of responses in specific domains without the need for extensive retraining contributes to meeting the growing demand for models for multilingual natural language processing and code generation.

## References

1. Tunstall, L., Beeching, E., Lambert, N., Rajani, N., Rasul, K., Belkada, Y., Huang, S., von Werra, L., Fourrier, C., Habib, N. and Sarrazin, N., 2023. Zephyr: Direct distillation of lm alignment. arXiv preprint arXiv:2310.16944.
2. Hu, E.J., Shen, Y., Wallis, P., Allen-Zhu, Z., Li, Y., Wang, S., Wang, L. and Chen, W., 2021. Lora: Low-rank adaptation of large language models. arXiv preprint arXiv:2106.09685.
3. Dettmers, T., Pagnoni, A., Holtzman, A. and Zettlemoyer, L., 2023. Qlora: Efficient finetuning of quantized llms. *arXiv preprint arXiv:2305.14314*.
4. C.B. Pronin, A.V. Volosova, A.V. Ostroukh, Yu.N. Strogov, V.V. Kurbatov, A.S. Umarova. MexIvanov/zephyr-python-ru / Hugging Face, 2023, Available at: https://huggingface.co/MexIvanov/zephyr-python-ru
5. Volosova, A.V. 2022, Tehnologii iskusstvennogo intellekta v ULS-sistemah, uchebnoe posobie dlja vuzov [*Artificial intelligence technologies in ULS-systems: a textbook for universities*] – Sankt-Peterburg: Lan, 2022. – 308 p.: ill.
6. HuggingFaceH4/CodeAlpaca_20K / Hugging Face, 2023, Available at: https://huggingface.co/datasets/HuggingFaceH4/CodeAlpaca_20K/.

7. Edward Beeching, Clémentine Fourrier, Nathan Habib, Sheon Han, Nathan Lambert, Nazneen Rajani, Omar Sanseviero, Lewis Tunstall, Thomas Wolf. Open LLM Leaderboard / Hugging Face, 2023, Available at: https://huggingface.co/spaces/HuggingFaceH4/open_llm_leaderboard
8. Pronin C.B., Volosova A.V., Ostroukh A.V., Strogov Y.N. Overcoming linguistic barriers in code assistants: creating a Qlora adapter to improve support for Russian-language coding instructions. (in Russian) // Dynamics of complex systems. V. 18, № 1, 2024, P. 32−40.

**Author Details:**

**Pronin Cesar B.** (b. 1999), Assistant at the Department of Automated Control Systems, Faculty of Management, Moscow Automobile and Road State Technical University (MADI). Email: caesarpr12@gmail.com

**Volosova Aleksandra V.** (b. 1971), Associate Professor, Candidate of Technical Sciences, Associate Professor of the Department of Information Processing and Management Systems, Faculty of Informatics and Management Systems, Bauman Moscow State Technical University. Email: volosova@bmstu.ru

**Ostroukh Andrey V. (**b. 1975), Professor, Doctor of Technical Sciences, Professor of the Department of Automated Control Systems of the Faculty of Management of the Moscow Automobile and Road State Technical University (MADI). Email: ostroukh@mail.ru

**Strogov Yuri Nikolaevich (**b. 2003) student of the Department of Automated Control Systems of the Faculty of Management of the Moscow Automobile and Road State Technical University (MADI). Email: zelkame@gmail.com